\documentclass[runningheads]{llncs}
\usepackage[T1]{fontenc}
\usepackage{graphicx}
\usepackage{array}
\usepackage{textcomp}
\usepackage{stfloats}
\usepackage{url}
\usepackage{verbatim}
\usepackage{color}
\usepackage{svg}
\usepackage{pdfpages}
\usepackage{pifont}
\usepackage{threeparttable}
\usepackage[linesnumbered,ruled,vlined]{algorithm2e}
\RequirePackage{letltxmacro}
\usepackage{comment}
\usepackage{paralist}
\LetLtxMacro{\LaTeXtextbf}{\textbf}
\LetLtxMacro{\textbf}{\LaTeXtextbf}
\usepackage[bookmarks=false]{hyperref} 
\usepackage{cite}
\hypersetup{
    colorlinks=true,
    citecolor=blue
}
\usepackage{amsmath,amssymb,amsfonts}
\usepackage{textcomp}

\usepackage{caption}
\usepackage{subcaption}
\usepackage{algpseudocode}
\usepackage{enumitem,lipsum}
\setlist[itemize,enumerate]{leftmargin=*}
\usepackage{etoolbox}
\makeatletter
\usepackage{wasysym}
\makeatother
\def\BibTeX{{\rm B\kern-.05em{\sc i\kern-.025em b}\kern-.08em
    T\kern-.1667em\lower.7ex\hbox{E}\kern-.125emX}}
\usepackage{tcolorbox}
\tcbuselibrary{breakable}
\usepackage{svg}
\usepackage{multirow}
\usepackage{listings}

\newcommand{\ie}{\textit{i.e.}}
\newcommand{\eg}{\textit{e.g.}}

\newlist{circlelist}{enumerate}{1}
\setlist[circlelist,1]{label=\textbf{\textcircled{\small\arabic*}}, leftmargin=*}

\usepackage{enumitem}
\usepackage{orcidlink}
\usepackage{booktabs}

\definecolor{dred}{RGB}{178,34,34}
\definecolor{dgreen}{RGB}{34,110,34}
\newcommand{\cmark}{\textcolor{dgreen}{\ding{51}}}
\newcommand{\xmark}{\textcolor{dred}{\ding{55}}}

\usepackage{tikz}
\usetikzlibrary{
arrows.meta,
positioning,
shapes,
fit,
calc
}

\begin{document}
%

%
%
\title{TACTIC-KG: Toward Small Agent Teams for Cyber Threat Intelligence Knowledge Graph Construction}

\author{Mouhamed Amine Bouchiha\inst{1}\orcidID{0000-0001-6142-6855} \and
Gregory Blanc\inst{1}\orcidID{0000-0001-8150-6617} 
}
\authorrunning{M. Bouchiha et al.}
%
\institute{SAMOVAR, Télécom SudParis, Institut Polytechnique de Paris 
\email{mbouchiha@telecom-sudparis.eu,} 
\email{gregory.blanc@telecom-sudparis.eu}
}

\maketitle              
\begin{abstract}
Cyber Threat Intelligence (CTI) reports are predominantly unstructured, heterogeneous, and noisy, which limits their direct usability for automated analysis and reasoning. Cybersecurity Knowledge Graphs (CSKGs) provide a structured representation of adversarial entities, actions, and relations, but constructing such graphs from free-text CTI remains a challenge. Recent approaches rely on monolithic Large Language Models (LLMs) to perform end-to-end extraction and completion, leading to high cost, limited controllability, and unstable performance. This paper introduces \textbf{TACTIC-KG}, an agentic framework for CSKG construction that decomposes the task into modular, specialized LLM agents responsible for extraction, typing, verification, and curation. Using lightweight models (3B--8B), TACTIC-KG improves stability, recall, and graph consistency while reducing deployment cost. We implement and evaluate TACTIC-KG against recent state-of-the-art systems. Experiments on human-annotated CTI reports show that agent specialization consistently outperforms larger monolithic in-context-learning (ICL) baselines in extraction F1-score, typing accuracy, and structural graph similarity.

\end{abstract}

\section{Introduction}
\label{sec:intro}

Cyber Threat Intelligence (CTI) reports contain rich textual narratives describing adversarial operations, tactics, and observable indicators. However, transforming these heterogeneous and unstructured reports into processable representations remains a fundamental challenge~\cite{buchel2025sok, zhao2024survey}. Traditional natural language processing techniques, such as named entity recognition (NER)~\cite{gao2021enabling} or classification-based pipelines~\cite{alam2023looking}, often fail to capture the nuanced and implicit relationships present in CTI, particularly for complex or fine-grained adversarial behaviors~\cite{buchel2025sok}.

Cybersecurity or CTI Knowledge Graphs (CSKGs) have emerged as a promising solution to structure CTI into entities, relationships, and contextual knowledge, supporting tasks such as visualization, attack-path reasoning, and automated correlation~\cite{cheng2025ctinexus, li2022attackg, zhao2024survey}. Recent research has leveraged Large Language Models (LLMs) to automate CSKG construction, taking advantage of their ability to process free-form text and capture latent semantic patterns. Prompt-based systems such as aCTIon~\cite{siracusano2023action} demonstrated the feasibility of LLM-driven document-level extraction of ATT\&CK techniques, although static prompts can suffer from hallucinations and limited precision on fine-grained entities. Ontology-grounded approaches, exemplified by CTINEXUS~\cite{cheng2025ctinexus} and IntelEX~\cite{xu2024intelex}, improve semantic grounding and entity canonicalization through in-context learning (ICL)~\cite{dong2024survey} and retrieval-augmented generation (RAG)~\cite{arslan2024survey}, respectively, but at the cost of increased system complexity and sensitivity to retrieval quality.

Complementary efforts focus on fine-tuning and domain adaptation of LLMs. Systems such as AECR~\cite{chen2025aecr} and Fengrui et al.~\cite{fengrui2024few} show that supervised and synthetic fine-tuning can enhance precision and reduce hallucinations, even with smaller model sizes. Comparative studies indicate that, while LLMs generally achieve higher recall than transformer-based classifiers, precision can remain a challenge, motivating hybrid or multi-model architectures~\cite{fayyazi2024advancing, kumarasinghe2024semantic}. Non-LLM approaches, including AttacKG+~\cite{li2022attackg}, LADDER~\cite{alam2023looking}, and CRUcialG~\cite{cheng2025crucialg}, highlight the continued relevance of structured pipelines and expert rules, although these methods can be brittle to linguistic variation.

In this work, we ask the question of whether constructing a CSKG from CTI should be treated as a \emph{multi-agent reasoning problem}, rather than a single monolithic inference task to minimize hallucination and improve stability. 

We introduce \textbf{TACTIC-KG}\footnote{Code and datasets: \href{https://github.com/mohaminemed/TACTIC-KG}{https://github.com/mohaminemed/TACTIC-KG}.}, an agentic framework that decomposes \textbf{extraction}, \textbf{typing}, \textbf{verification}, and \textbf{curation} into specialized small LLM agents. By combining multi-agent orchestration with ontology grounding, TACTIC-KG improves robustness and stability, while enabling the use of lightweight models. Unlike static or monolithic LLM approaches, our framework explicitly manages uncertainty, enforces ontology compliance to construct accurate and actionable CSKGs from unstructured CTI reports.

The remainder of this paper is structured as follows. Sect.~\ref{sec:relatedwork} reviews related work. Sect.~\ref{sec:problemdef} formalizes the CSKG construction problem. Sect.~\ref{sec:framework} presents the TACTIC-KG framework. Sect.~\ref{sec:experiments} describes the experimental setup and evaluation. Sect.~\ref{sec:discussion} discusses insights and limitations, and Sect.~\ref{sec:conclusion} concludes the paper.

\begin{table}[t]
\centering
\caption{Comparison of cybersecurity IE and CSKG construction properties.}
\label{tab:related_work}
\resizebox{\linewidth}{!}{
\begin{tabular}{lcccccc}
\toprule
\textbf{Property} & \textbf{CTINEXUS~\cite{cheng2025ctinexus}} & \textbf{IntelEX~\cite{xu2024intelex}} & \textbf{AECR~\cite{chen2025aecr}} & \textbf{CRUcialG~\cite{cheng2025crucialg}} & \textbf{AttacKG~\cite{li2022attackg}} & \textbf{TACTIC-KG} \\
\midrule
Faithfulness verification      & \xmark & $\triangle$ & $\triangle$ & \xmark & \xmark & \cmark \\
Hallucination mitigation       & $\triangle$ & $\triangle$ & \cmark & $\triangle$ & \xmark & \cmark \\
Over-completion control        & \xmark & \xmark & \xmark & \xmark & \xmark & \cmark \\
Ontology compliance            & \cmark & $\triangle$ & \xmark & \cmark & \cmark & \cmark \\
Structural consistency         & \cmark & \xmark & $\triangle$ & $\triangle$ & \cmark & \cmark \\
Small-model compatibility      & $\triangle$ & \cmark & \cmark & \cmark & \cmark & \cmark \\
Incomplete-evidence handling   & \cmark & \xmark & \xmark & $\triangle$ & $\triangle$ & \cmark \\
\bottomrule
\end{tabular}}
\begin{flushleft}
	\scriptsize
	\cmark: Explicitly supported \quad
	$\triangle$: Partially supported \quad
	\xmark: Not addressed
\end{flushleft}
\end{table}

\section{Related work}
\label{sec:relatedwork}
This section reviews relevant studies on cybersecurity information extraction (IE \eg, TTPs) and CSKGs construction from unstructured CTI reports. Table~\ref{tab:related_work} compares TACTIC-KG with the most recent studies.

\subsubsection{Ontology-Grounded \& Prompting LLM Systems.}
aCTIon~\cite{siracusano2023action} represents one of the earliest LLM-based systems, relying on GPT-3.5-Turbo with static prompts to perform document-level summarization and ATT\&CK technique extraction. While flexible and easy to deploy, static prompting strategies suffer from limited precision and weak control over hallucinations, particularly for fine-grained techniques. CTINEXUS~\cite{cheng2025ctinexus} introduces a three-phase pipeline for constructing CSKGs using LLMs through ICL~\cite{dong2024survey}. The workflow integrates triplet extraction from CTI reports, hierarchical entity alignment, and long-distance relation inference to merge disconnected subgraphs. The system uses a $k$-Nearest Neighbors (kNN) retriever for dynamic ICL demo~\cite{zhao2024knn} 
selection, executes a single-inference triplet extraction prompt, and combines LLM-based typing with embedding-based merging to produce CSKG graphs aligned with the the MALOnt ontology~\cite{rastogi2020malont}. CTINEXUS provides improved semantic grounding compared to static prompting, but notes increased system complexity and sensitivity to retrieval quality. Similarly, IntelEX~\cite{xu2024intelex} adopts GPT-4o-mini with RAG and a multi-agent design to improve precision. While effective for extraction, its performance remains dependent on the accuracy of the retrieved context.

\subsubsection{Fine-Tuned and Domain-Adaptive LLMs.}
AECR~\cite{chen2025aecr} explores fine-tuning ChatGLM3-6B using supervised fine-tuning combined with a classification head. This approach outperforms larger general-purpose LLMs in precision while significantly reducing hallucinations, illustrating the benefits of domain adaptation despite model size constraints. Fengrui et al.~\cite{fengrui2024few} fine-tune Llama2-7B using GPT-4–augmented training data, demonstrating that synthetic supervision can partially compensate for the scarcity of expert-labeled CTI datasets. The key advantage of fine-tuning  lies in its reduced dependence on prompt design, as the model learns during training what information should be extracted from the input. Fieblinger et al.~\cite{fieblinger2024actionable} further explore the use of open-source LLMs (\eg, Llama 2, Mistral 7B Instruct, and Zephyr) for extracting structured 
triplets from CTI reports and constructing KGs. Their study compares prompt engineering, guided generation, and fine-tuning, showing that fine-tuning outperforms prompt-based approaches in extraction quality. However, they also highlight persistent challenges in scaling LLM-based pipelines for large-scale KG construction and downstream tasks (\eg, link prediction).

\subsubsection{Non-LLM Recent Approaches.}
Beyond LLM-based extraction, AttacKG~\cite{li2022attackg} proposes a multi-stage pipeline that combines TTP extraction with knowledge graph construction, forming an end-to-end CTI structuring system rather than a pure extraction approach. In a similar vein, LADDER~\cite{alam2023looking} introduces a scalable framework for extracting text-based attack patterns from CTI reports by modeling attack phases across Android and enterprise environments, and mapping them to the MITRE ATT\&CK framework. CRUcialG~\cite{cheng2025crucialg} further extends this direction with a unified framework for reconstructing integrated and rational attack graphs, targeting connectivity, rationality, and universality. However, these approaches largely depend on predefined schemas, pattern matching, or manually curated expert rules. As a result, they are brittle to linguistic variation: paraphrasing, implicit relations, or diverse narrative styles in CTI reports can break rule-based matching and lead to missed or incorrect extractions. Additionally, temporal order is often inferred from textual sequence, which is unreliable since CTI reports frequently describe events in a non-chronological manner.

\subsubsection{Cyber Threat Ontologies.} A variety of ontologies have been proposed by both academia and industry to describe adversarial actions and cyber threat intelligence. Among them, STIX~2.1~\cite{stix21} has been widely adopted as a standard representation format, as it provides a generic, extensible, and interoperable framework for modeling threat intelligence. As a result, many organizations and platforms rely on STIX to exchange and operationalize CTI data. Beyond this generic representation layer, several more specialized models and
ontologies have been developed, often encoded using STIX~2.1. Notable examples include the MISP~\cite{misp} data model for representing threat objects and galaxies, as well as MITRE frameworks such as ATT\&CK~\cite{mitre2023}, D3FEND, CAPEC, CWE, and CVE. These frameworks differ in scope and intent, covering defensive techniques, attack patterns, software weaknesses, and known vulnerabilities. Among these, the ATT\&CK framework has emerged as the de facto standard for describing adversary Tactics, Techniques, and Procedures (TTPs). 
MALOnt~\cite{rastogi2020malont} is another comprehensive ontology that defines 33 entity types (including 17 primary types and 16 sub-types) and 27 relation types, enabling rich and expressive knowledge modeling. It covers a wide spectrum of cyber threat intelligence entities (\eg, \emph{Account}, \emph{Threat Actor}, \emph{Event}, ...). Furthermore, MALOnt’s schema is well aligned with natural language descriptions commonly found in cyber threat reports. This makes it suitable for automated knowledge extraction pipelines~\cite{cheng2025ctinexus, waldrop2026semantic}.

\begin{figure}[t]
\centering
\hspace{-0.1\linewidth}
\resizebox{0.38\linewidth}{!}{
\begin{subfigure}[t]{0.58\linewidth}
\centering
\begin{tikzpicture}[
    entity/.style={circle, draw, minimum size=1cm, text centered, font=\scriptsize},
    entityo/.style={circle, draw, dashed,  minimum size=1cm, text centered, font=\scriptsize},
    correct/.style={fill=green!30},
    partial/.style={fill=yellow!30},
    missing/.style={fill=red!30},
    relation/.style={->, thick, >=stealth},
    other/.style={->, thick, dashed, >=stealth} 
]

\node[entity, correct, align=center] (ios) at (0,0) {iOS 14.8 \\ Tool};
\node[entity, correct, align=center] (cve) at (2.8,0) {FORCEDENTRY \\ Vulnerability};
\node[entity, correct, align=center] (pegasus) at (6.9,0) {Pegasus \\ Malware};
\node[entity, partial, align=center] (group) at (3,-3) {NSO Group \\ Organization};
\node[entity, missing, align=center] (journalist) at (6.9,-3) {Journalists \\ Unknown};
\node[entity, missing, align=center] (activist) at (5,-3) {Activists \\ Unknown};
\node[entityo, align=center] (other) at (0.3,-2) {entity \\ type};

\draw[relation] (ios) -- node[above]{fixes} (cve);
\draw[relation] (cve) -- node[above]{used to deploy} (pegasus);
\draw[relation] (pegasus) -- node[right]{targets} (journalist);
\draw[relation] (pegasus) -- node[left]{targets} (activist);

\draw[other] (cve) -- node[above]{\small relation} (other);
\draw[other] (group) -- node[above]{\small reported on} (other);

\node[anchor=south] at (4,1.5) {\textbf{CTINEXUS}};

\end{tikzpicture}
\caption{CTINEXUS: Deepseek-v3.1-671B + ICL}
\label{fig:ctinexus_kg}
\end{subfigure}}
\hspace{0.1\linewidth}
\resizebox{0.38\linewidth}{!}{
\begin{subfigure}[t]{0.58\linewidth}
\centering
\begin{tikzpicture}[
    entity/.style={circle, draw, minimum size=1cm, text centered, font=\scriptsize},
    entityo/.style={circle, draw, dashed,  minimum size=1cm, text centered, font=\scriptsize},
    correct/.style={fill=green!30},
    partial/.style={fill=yellow!30},
    missing/.style={fill=red!30},
    relation/.style={->, thick, >=stealth},
     other/.style={->, thick, dashed, >=stealth} 
]

\node[entity, correct, align=center] (ios2) at (0,0) {iOS 14.8 \\ Tool};
\node[entity, correct, align=center] (cve2) at (2.8,0) {FORCEDENTRY \\ Vulnerability};
\node[entity, correct, align=center] (pegasus2) at (6.9,0) {Pegasus \\ Malware};
\node[entity, correct, align=center] (journalist2) at (6.9,-3) {Journalists, \\ activists,.. \\ Organization};
\node[entity, correct, align=center] (attacker2) at (2.8,-3) {An attacker \\ Attacker};
\node[entity, correct, align=center] (appleid2) at (0,-3) {Apple ID \\ Credential};
\node[entity, correct, align=center] (time2) at (5,-3) {Feb 2021 \\ Time};
\node[entityo, align=center] (other2) at (0.3,-1.45) {entity \\ type};

\draw[relation] (ios2) -- node[above]{fixes} (cve2);
\draw[relation] (cve2) -- node[above]{used to deploy} (pegasus2);
\draw[relation] (pegasus2) -- node[left]{targets} (journalist2);
\draw[relation] (attacker2) -- node[left]{exploits} (cve2);
\draw[relation] (attacker2) -- node[above]{requires} (appleid2);
\draw[relation] (cve2) -- node[right]{used since} (time2);

\draw[other] (cve2) -- node[above]{\small relation} (other2);

\node[anchor=south] at (4,1.5) {\textbf{TACTIC-KG}};

\end{tikzpicture}
\caption{TACTIC-KG: Ministral-3-3B + LoRA}
\label{fig:tactic_kg}
\end{subfigure}}

\begin{tikzpicture}[
     report/.style={rectangle, draw, dashed, rounded corners, fill=gray!15, text centered, text width=11cm}
]
\node[report] at (0,1.5) {\tiny \textbf{Input report:} ``\textit{The recent \textbf{iOS 14.8 update} fixes a zero-day, zero-click exploit for a vulne\-rability affecting every mobile iOS device. The flaw, dubbed \textbf{FORCEDENTRY} (CVE-2021-30860), resided in Apple's iMessage and, according to a report by The Citizen Lab, was used to push NSO Group's \textbf{Pegasus} spyware to mobile iOS devices dating back to as far as \textbf{February 2021}. \dots the spyware targeting \textbf{journalists, activists}, and others \dots \textbf{An attacker} exploiting the flaw only needs the \textbf{Apple ID} of a device in order to silently compromise it.''}};
\end{tikzpicture}

\caption{CSKG construction comparison for a real-world CTI report: 
\textcolor{green}{\textbf{Green}} nodes = correctly typed; \textcolor{yellow}{\textbf{yellow}} = partially correct; \textcolor{red}{\textbf{red}} = missing or mis-typed. 
CTINEXUS misses attacker, temporal, and some entity typing; TACTIC-KG captures all entities.} 
\label{fig:kg_comparison}
\end{figure}
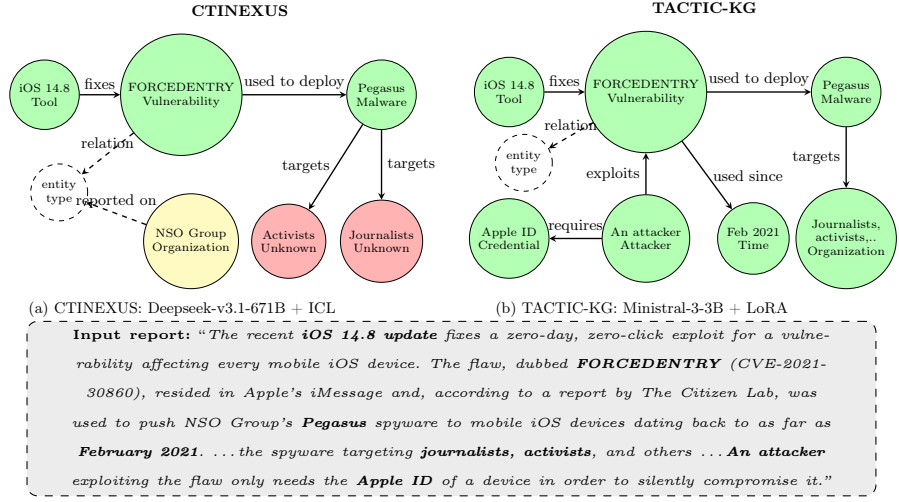

\section{Problem Definition}
\label{sec:problemdef}

Let $R$ denote an unstructured CTI report in free-form natural language. The objective of CSKG construction is to transform $R$ into a structured graph 
\[
G = (V, E),
\]
where $V$ is a set of typed entities and $E$ is a set of typed relations between entities, subject to ontology constraints $\mathcal{O}$ (\eg, MALOnt~\cite{rastogi2020malont}, ATT\&CK~\cite{mitre2023}, STIX~\cite{stix21}). Formally, the construction process consists of three coupled tasks:

\begin{enumerate}
    \item \textbf{Extraction:} Identify candidate triples $(h, r, t)$ such that $h,t \in V$ and $r \in \mathcal{R}$ are supported by textual evidence in $R$.
    \item \textbf{Typing:} Assign ontology-compliant semantic types to $h$, $r$, and $t$.
    \item \textbf{Graph Assembly:} Integrate validated triples into a consistent graph $G$.
\end{enumerate}

While this formulation appears straightforward, constructing $G$ from $R$ is inherently challenging due to several interacting sources of uncertainty. As illustrated in Fig.~\ref{fig:kg_comparison}, even when powered by a strong LLM such as DeepSeek-V3.1-671B, CTINEXUS~\cite{cheng2025ctinexus} fails to capture key elements, including attacker roles, temporal information, and parts of target typing, whereas TACTIC-KG extracts and types all entities correctly. Moreover, as we demonstrate in Sect.~\ref{sec:experiments}, CTINEXUS exhibits significantly degraded CSKG quality when using smaller models. We detail the underlying sources of this uncertainty and instability in the following.
 
\subsubsection{Hallucination and Faithfulness.} \label{subsec:faithfulness}

LLMs may generate triples that are \emph{plausible} but not \emph{entailed} by $R$. We define a triple $(h,r,t)$ as \emph{faithful} if and only if it is semantically supported by explicit or implicit evidence in $R$. Otherwise, it is considered a \emph{hallucinated triple}. Hallucinations arise from \begin{inparaenum}[(i)]
    \item \textit{domain knowledge gaps:} the model lacks prior exposure to specific malware families, vulnerabilities, or niche TTP patterns;
    \item \textit{insufficient context:} retrieval errors, truncated or incomplete passages (\eg, naive chunking) lead to unsupported inference; or
    \item \textit{context overload:} when excessive or irrelevant tokens dilute attention, critical evidence loses representational weight, causing entity substitutions or fabricated relations.
\end{inparaenum} Unlike traditional extraction errors, hallucinated triples may appear syntactically and semantically coherent, making them difficult to detect through 
automatic
validation. In a pipeline setting, such errors propagate and may distort the final graph structure.

\subsubsection{Over-Completion and Artificial Connectivity.} \label{subsec:overcopletion}

Many CSKG construction systems~\cite{cheng2025ctinexus, cheng2025crucialg} implicitly favor the production of connected graphs. However, CTI reports do not necessarily describe fully connected structures (\eg, complete attack chains). Human-annotated datasets\cite{della2025cti} frequently contain disconnected components, reflecting partial observations or incomplete forensic evidence. We therefore distinguish between 
\begin{inparaenum}[(i)]
    \item \textit{evidence-grounded inference:} relations supported by textual or logically entailed evidence; and
    \item \textit{graph over-completion:} artificially introducing edges to increase connectivity without textual support.
\end{inparaenum} 
Over-completion is a structural manifestation of hallucination: the system prioritizes graph connection over faithfulness.

\subsubsection{Context-Size Trade-Off.} \label{subsec:context-size}

Let $\mathcal{C}(R)$ denote the context window provided to the LLM. There exists a non-trivial trade-off: 
\begin{inparaenum}[(i)]
    \item small $\mathcal{C}(R)$ increases omission errors due to missing evidence; and
    \item large $\mathcal{C}(R)$ increases attention diffusion, reducing signal-to-noise ratio and increasing hallucination risk.
\end{inparaenum}
This trade-off is particularly acute for lightweight models with limited capacity. While techniques such as ICL~\cite{dong2024survey}, RAG~\cite{arslan2024survey}, or even dynamic context construction~\cite{zhang2025agentic} partially mitigate this issue for large models, their effectiveness largely depends on the underlying model capacity and the retrieval quality. In practice, these techniques provide substantial gains for large, high-capacity models, but offer limited benefits for smaller models, where constrained reasoning and context hinder their impact~\cite{agarwal2024many, wei2023larger}.

\subsubsection{Problem Statement.}
Given an unstructured CTI report $R$ and ontology constraints $\mathcal{O}$, construct a knowledge graph $G = (V,E)$ that satisfies the following constraints:
\begin{center}
\begin{tabular}{c|c}
\textbf{Faithfulness} \quad & $\forall (h,r,t) \in E, \; (h,r,t) \models R $\\
\textbf{Ontology Compliance} \quad & $\text{types}(h,r,t) \in \mathcal{O}$ \\
\textbf{Minimal Over-Completion} \quad & $E \text{ contains no unsupported edges}$ \\
\textbf{Structural Consistency} \quad & $G \text{ satisfies global logical constraints}$
\end{tabular}\\
\end{center}
In other words, the constructed graph must remain fully grounded in the input report, respect ontology constraints, and avoid introducing spurious relations. The central research question of this work is therefore:
\begin{quote}
\emph{Can CSKG construction be reformulated as a controlled multi-agent reasoning process that verifies faithfulness and limits hallucination propagation, rather than relying on monolithic generation or connectivity-driven inference?}
\end{quote}

\section{TACTIC-KG Framework}
\label{sec:framework}

TACTIC-KG reformulates CSKG construction as a controlled multi-stage reasoning process. Instead of relying on a single monolithic LLM to jointly extract, type, and assemble the graph, we decompose the problem into specialized agents operating under constrained objectives. The core design principle is \textit{faithfulness before connectivity}. 
Each stage is responsible for a narrowly defined transformation, and agents are not allowed to introduce information not grounded in the input report. This modularization 
reduces hallucination propagation and enables fine-grained supervision. The decomposition yields three robustness advantages:

\begin{enumerate}
    \item \textbf{Reduced hallucination surface:} 
    Each agent operates within a constrained reasoning space.
    \item \textbf{Error containment:} 
    Hallucinated triples do not automatically propagate.
    \item \textbf{Task entropy reduction:} 
    Smaller models perform better when solving narrowly scoped tasks.
\end{enumerate}

Empirically (Sect.~\ref{sec:experiments}), we observe that even strong large-scale models can hallucinate 
entity substitutions or unsupported relations. 
In monolithic pipelines, such errors propagate and distort graph structure. 
TACTIC-KG localizes these risks and enforces faithfulness as a first-class constraint.

\begin{figure}[t]
\centering
\includegraphics[width=\linewidth]{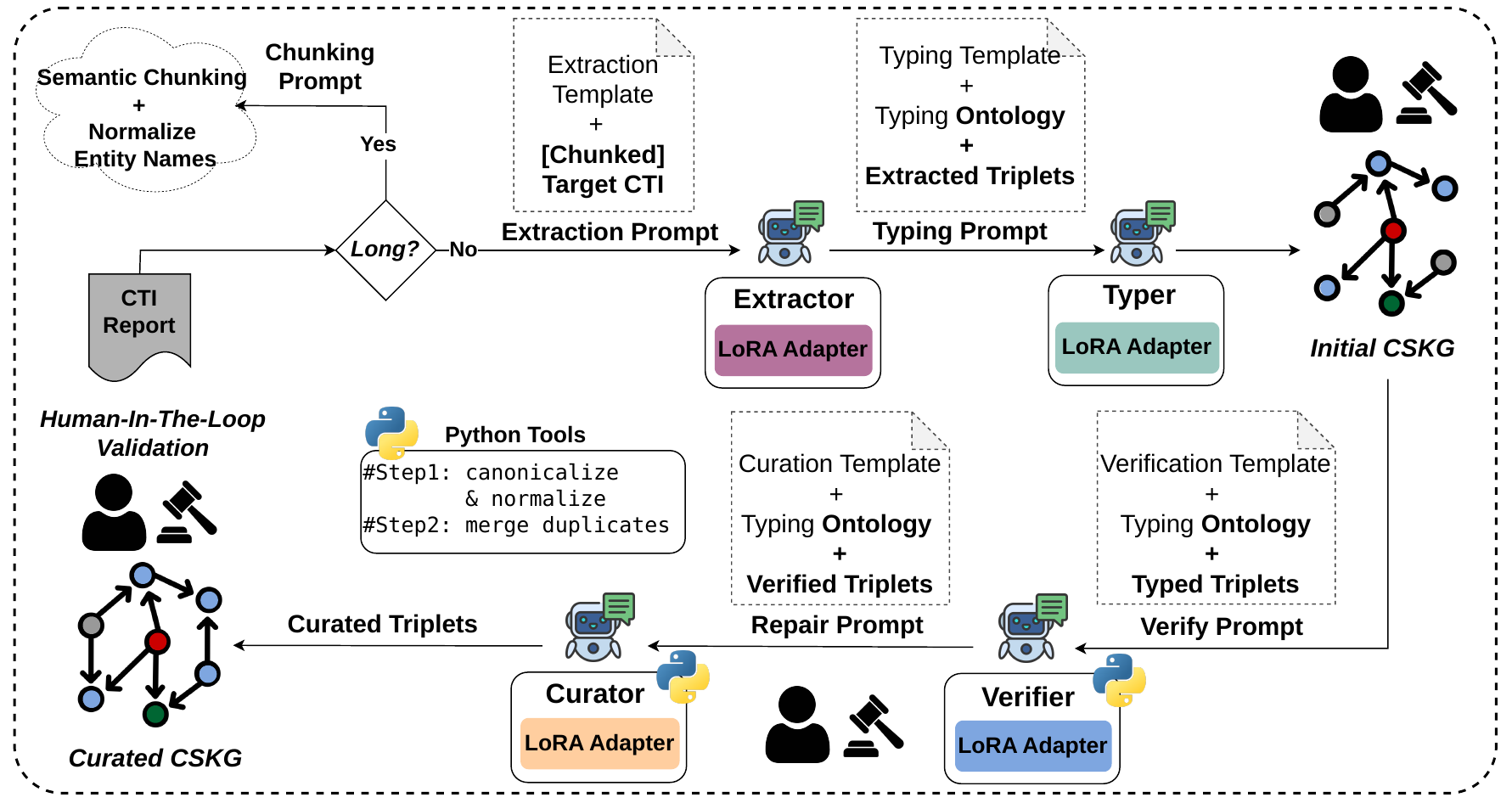}
\caption{TACTIC-KG framework.}
\label{fig:TACTIC-KG}
\end{figure}

\subsection{High-Level Workflow}
\label{subsec:workflow}

As shown in Fig.~\ref{fig:TACTIC-KG}, given a CTI report $R$, the system first applies \textit{semantic chunking} to handle long-form documents that exceed model context limits. 
Rather than naïve fixed-length segmentation, chunking is performed at sentence or discourse boundaries with controlled overlap, preserving local semantic coherence while preventing context dilution. The resulting set of chunked reports $\{R_1, R_2, ..., R_n\}$ is processed sequentially through a multi-agent pipeline:

\begin{center}
\texttt{Raw Report}
$\rightarrow$
\textbf{Semantic Chunking}
$\rightarrow$
\texttt{Chunked Reports}
$\rightarrow$
\textbf{Extractor}
$\rightarrow$
\textbf{Typer}
$\rightarrow$
\texttt{Initial CSKG}
$\rightarrow$
\textbf{Verifier}
$\rightarrow$
\textbf{Curator}
$\rightarrow$
\texttt{Curated CSKG}
\end{center}

A long raw CTI report is first segmented using semantic chunking to preserve discourse boundaries. Next, the agents interact sequentially under an auditable and Human-in-the-loop (HITL)-friendly protocol such that: \begin{inparaenum}[(i)]
    \item intermediate outputs are serialized in JSON format,
    \item no agent can modify upstream spans,
    and \item partial re-execution is supported.
\end{inparaenum} In this setting, the pipeline includes:
\begin{itemize}
    \item \textbf{Extractor Agent:} 
    Generates candidate triples $(h,r,t)$ grounded in each chunk $R_i$. No typing or global reasoning is performed at this stage.

    \item \textbf{Typer Agent:} 
    Assigns ontology-compliant types $(\tau_h, \tau_t)$ to entities using local context and relation semantics.
    \item \textbf{Verifier Agent:} 
    Operates at the triplet level. Removes unsupported or low-confidence triples by structural and ontology validation.
    \item \textbf{Curator Agent:} 
    Operates at the document level after merging verification outputs. Introduces only logically necessary structural edges (\eg, normalization links, alias resolution: ``TrickBot malware and TrickBot'').

\end{itemize}

\subsection{Faithful Span-Level Generation: Extractor Agent}

The Extractor Agent is responsible solely for identifying candidate triples 
$(h, r, t)$ from raw CTI text. It does \emph{not} perform typing or canonicalization. This separation prevents premature abstraction, which is a major source of hallucinated structure in end-to-end systems.

\textit{Confidence estimation.} After generation, token-level log-likelihood scores are computed. Triple-level confidence is derived via aggregation and propagated downstream. Low-confidence triples can later be flagged by the verifier.

\textit{Context management.} To mitigate context overload (Sect.~\ref{subsec:faithfulness}), long reports are segmented using sentence-level chunking with controlled overlap. Triples are generated per chunk and later merged. This reduces attention dilution while preserving local semantic coherence.

\subsection{Ontology-Constrained Classification: Typer Agent}
Joint extraction and typing forces the Typer Agent model to reason over both span detection 
and ontology inference simultaneously, increasing cognitive load and hallucination risk. Decoupling reduces task entropy and allows targeted fine-tuning. The Typer Agent assigns semantic types to extracted entities using: \begin{inparaenum}[(i)]
    \item local textual context,
    \item relation semantics,
    \item and a fixed ontology schema.
\end{inparaenum} Formally, for each extracted triple $(h,r,t)$, the typer predicts:
\[
(h, r, t) \rightarrow (h, \tau_h, r, t, \tau_t)
\]
where $\tau_h, \tau_t \in \mathcal{O}$. The typer is restricted to a predefined class set. If the evidence is insufficient, it assigns \texttt{Uncategorized}, which prevents type ontology drift.

\subsection{Evidence-Grounded Triplet Validation: Verifier Agent}

The Verifier Agent acts as a filtering stage between semantic typing and graph curation. Its role is to ensure that every candidate triplet is \emph{faithfully grounded in the source text}, thereby preventing unsupported or hallucinated relations from propagating downstream. Formally, given a typed triplet $(h, \tau_h, r, t, \tau_t)$ and an input text $x$, the verifier performs a binary classification:
\[
\mathcal{V}(h, \tau_h, r, t, \tau_t, x) \rightarrow \{ \texttt{SUPPORTED}, \texttt{NOT\_SUPPORTED} \}
\]

A triplet is labeled \texttt{SUPPORTED} if the relation between $h$ and $t$ is either explicitly stated or implicitly entailed in $x$. 
We define \begin{inparaenum}[(i)] \item \textit{explicit support} as cases where the relation is directly expressed in the text via a lexical or syntactic realization of the predicate (or its paraphrase). For example, in the sentence \textit{``The malware Emotet communicates with its C2 server at domain X''}, the triplet (Emotet, communicates\_with, C2 server) is explicitly supported.

\item \textit{Implicit support} refers to cases where the relation is not directly stated but can be inferred from context or domain knowledge. For example, from \textit{``The attacker used TrickBot to deploy ransomware on the victim system''}, we infer (TrickBot, enables, ransomware deployment), even though the enabling relation is not explicitly expressed. \end{inparaenum} Otherwise, the triplet is labeled \texttt{NOT\_SUPPORTED}.

\textit{Evidence extraction.}
Beyond classification, the verifier enhances interpretability by extracting a minimal supporting sentence $e \subset x$ whenever possible, providing a verifiable trace for downstream HITL validation:
\[
\mathcal{V}_e(h, r, t, x) \rightarrow e \;\; \text{or} \;\; \varnothing
\]

\textit{Strict grounding constraints.}
The verifier operates under a closed-world assumption with respect to the input text: \begin{inparaenum}[(i)]
    \item only the provided text $x$ may be used as evidence,
    \item external knowledge is forbidden,
    \item plausible but unstated relations must be rejected.
\end{inparaenum} By explicitly enforcing textual entailment at the triplet level, the verifier filters hallucinations introduced during extraction and typing. This transforms the pipeline into a \emph{generate-then-verify} paradigm, where unsupported structures are pruned before reaching the curator.

\subsection{Structural Completion and Sanitization: Curator Agent}

The Curator Agent performs the final sanitization pass before graph insertion. Its responsibilities include: \begin{inparaenum}[(i)]
    \item deduplication and canonicalization,
    \item entity normalization (case, formatting, alias merging),
    \item ontology compliance enforcement,
    \item graph-level constraint validation.
 \end{inparaenum}The curator does not perform unconstrained generation. 
Instead, it operates under a strictly bounded repair objective: 
it may introduce new triplets only when they are supported by textual evidence. In addition to constrained completion, the curator acts as a structural firewall 
by filtering unsupported or low-confidence triplets propagated from upstream agents. Thus, it prevents hallucinated or inconsistent edges from entering the final graph. This dual role—verification and minimal constrained repair—preserves global graph coherence.

\subsection{Parameter-Efficient Specialization via LoRA}

To enable efficient specialization without full-model fine-tuning, we employ Low-Rank Adaptation (LoRA)~\cite{hu2022lora} on lightweight 3B--8B parameter models. Large frontier models reduce hallucination, but incur prohibitive cost and latency~\cite{cheng2025ctinexus}. 
Instead, we adapt smaller instruction-tuned models using LoRA to specialize them for structured JSON generation, faithful span extraction, and ontology-constrained typing. LoRA offers~\cite{shuttleworth2024lora}: \begin{inparaenum}[(i)]
    \item reduced GPU memory footprint,
    \item modular agent-specific adapters, and
    \item independent upgrading of agents.
\end{inparaenum}

\quad \textbf{Extractor Fine-Tuning.} The extractor is trained using:
\begin{inparaenum}[(i)]
    \item prompt-target concatenation,
    \item masked\footnote{During training, the model sees both the input prompt and the expected output. But we do not want the model to learn to “predict” the prompt itself, only the answer.} prompt tokens,
    \item chunked reports with overlap,
    \item strict JSON-only output format.
\end{inparaenum} LoRA adapters are injected into attention projection layers 
(\eg, $q\_proj$, $v\_proj$), which allows task adaptation 
without modifying the base model weights.

\quad \textbf{Typer Fine-Tuning.} The typer is trained under a closed label space. The ontology is embedded directly in the prompt instruction. 
This effectively transforms typing into constrained structured generation 
rather than open-ended inference.

\quad \textbf{Verifier Fine-Tuning.} 
The verifier is trained as a triplet-level binary classifier under a strict evidence-grounding constraint. 
Each training instance consists of (i) the source text $x$, and (ii) a single candidate triplet $(h, r, t)$. 
The model is optimized to generate a structured JSON output containing the label 
$\{\texttt{SUPPORTED}, \texttt{NOT\_SUPPORTED}\}$. To ensure precise supervision and reduce ambiguity, training is performed at the \emph{single-triplet level}. The input is formatted as a constrained prompt embedding the text and the triplet, while the target output is the corresponding labeled JSON structure. The model is trained to rely on the provided text. This transforms verification into a textual entailment under strict grounding. Supervision is derived automatically from annotated CTI corpora. 
Positive samples correspond to supported triplets,  while negative samples are generated from non-supported or weakly implied relations. This construction aims at improving robustness against over-acceptance bias.

\quad \textbf{Curator Fine-Tuning.} The curator is fine-tuned to perform constrained graph repair.
Given (i) the source text, (ii) the current graph, the model is trained to predict the \emph{minimal set of triplets}
that complete the graph. Training targets are formed automatically. For each report, explicit triplets are partitioned into components. Ground-truth bridging triplets are derived from implicit (latent) relations that connect nodes across components. Only triplets that introduce cross-component edges are used as supervision. The model is required to output valid JSON containing only new bridging triplets. Loss is masked over the prompt portion (\ie, optimization applies only to the predicted repair edges). The curator is trained under a \emph{minimality objective}:
it must introduce the smallest set of structurally necessary connections
without repeating existing edges or hallucinating unsupported relations. This formulation transforms graph completion from open-ended generation
into constrained repair grounded in textual evidence.

\section{Experimental Evaluation}
\label{sec:experiments}
We evaluate TACTIC-KG on human-annotated CTI reports using semantic similarity matching~\cite{lairgi2026atom, liu2022unsupervised}. Metrics include extraction precision, recall, F1-score, typing accuracy~\cite{cheng2025ctinexus}, and graph structural similarity~\cite{liu2022unsupervised}. Baselines include multiple CTINEXUS~\cite{cheng2025ctinexus} configurations.  We structure our analysis around a set of research questions that investigate the effectiveness of TACTIC-KG across extraction quality, typing accuracy, and graph-level consistency. Our evaluation addresses the following research questions:

\begin{itemize}
    \item \textbf{RQ1:} Does agentic decomposition improve extraction quality?
    \item \textbf{RQ2:} How does agent specialization affect precision-recall (PR) trade-offs?
    \item \textbf{RQ3:} Does TACTIC-KG improve typing and semantic grounding and produce more structurally consistent graphs?
    \item \textbf{RQ4:} What is the impact of a hybrid agentic design?
    \item \textbf{RQ5:} How much data does TACTIC-KG need to reach monolithic model performance?
\end{itemize}

\subsection{Experimental Setup}

\textbf{Datasets.}
We construct a comprehensive CTI benchmark using multiple recent human-annotated resources. In particular, we built our dataset using two complementary corpora: (i) CTI-HAL~\cite{della2025cti}, a human-annotated CTI dataset, and (ii) the CTINEXUS benchmark~\cite{cheng2025ctinexus}, which provides fine-grained annotations for triplet extraction, entity alignment, and prediction tasks. CTI-HAL represents a large collection of CTI reports from multiple APT groups and cybercrime campaigns. The data includes detailed technical reports from well-known security vendors and threat intelligence platforms, spanning several prominent threat actors such as APT29, CARBANAK, FIN6, FIN7, OILRIG (APT34), SANDWORM, and WIZARD SPIDER. Each report contains rich unstructured descriptions of TTPs, malware families, infrastructure, and victimology. The aggregated corpus consists of over 230 human-annotated CTI reports. 

\textbf{Ontology.} As in~\cite{cheng2025ctinexus}, we use MALOnt~\cite{rastogi2020malont} for this evaluation, as it represents one of the most comprehensive open-source ontologies available. MALOnt provides fine-grained sub-typing, particularly for \emph{Indicator} and \emph{Malware Characteristics}, allowing for more precise semantic grounding. We use the revised version of MALOnt\footnote{\href{https://github.com/aiforsec/MALOnt}{https://github.com/aiforsec/MALOnt}} that is aligned with STIX~2.1, which further improves its practical applicability for structured threat intelligence extraction.

\medskip

\noindent\textbf{Annotation and Task Design.}
The CTINEXUS dataset provides annotations for triplet extraction, entity typing, and inter-entity relations. We ensure consistency across datasets via entity normalization (alias resolution and canonical mapping) and relation alignment to a unified CTINEXUS schema. For CTI-HAL reports, we apply distant supervision using ontology-guided pattern matching and threat-intelligence lexicons to generate noisy candidate triplets. These are then refined through manual curation, where a human annotator validates entities, corrects entity linking errors, and removes noisy relations.

\medskip
\noindent\textbf{Data Splits and Usage.}
We adopt the following (default) task-specific data partitioning strategy for our multi-agent pipeline:
\begin{itemize}
    \item \textbf{Extractor \& Typer Training:} up to 187 reports (539 chunks) are used to fine-tune the information extraction and entity typing modules.
    \item \textbf{Verifier \& Curator Training:} 85 reports are used for training verification and curation components.
    \item \textbf{Evaluation:} Three held-out test sets are constructed: TEST0 with the 11 test reports from~\cite{cheng2025ctinexus},  TEST1 with 25 reports and TEST2 with 45 reports, sampled to ensure diversity in report styles and extraction difficulty. TEST2 extends TEST1 with more challenging reports.
\end{itemize}

\textbf{Baselines.}
We compare TACTIC-KG against CTINEXUS~\cite{cheng2025ctinexus}, which introduces a three-phase pipeline for constructing CSKGs using LLMs through ICL~\cite{dong2024survey}. The workflow integrates (1) triplet extraction from CTI reports, (2) hierarchical entity alignment and canonicalization, and (3) long-distance relation inference to merge disconnected subgraphs. It uses a $k$-Nearest Neighbor (kNN) retriever for dynamic demonstration selection, executes a single-inference triplet extraction prompt, and combines LLM-based typing with embedding-based merging to produce connected graphs aligned with MALOnt ontology~\cite{rastogi2020malont}.

We evaluate a diverse set of recent LLMs, including large-scale monolithic models (\eg, DeepSeek-V3.1-671B, Kimi-K2-1T) and other open-weight models (\eg, GPT-OSS-20B). To this end, we adapt the CTINEXUS\footnote{\href{https://github.com/peng-gao-lab/CTINexus}{https://github.com/peng-gao-lab/CTINexus}} benchmark to support recent LLMs deployed via Ollama Cloud\footnote{\href{https://docs.ollama.com/cloud}{https://docs.ollama.com/cloud}}. The CTINEXUS baselines follow a single-model paradigm, where extraction, typing and link prediction are performed jointly in a monolithic pipeline.

\textbf{TACTIC-KG Configuration.}
As presented in Sect.~\ref{sec:framework}, TACTIC-KG decomposes the pipeline into four specialized agents. In our default setup, only extraction and typing are agent-specific, while verification and curation rely on a \textit{single fine-tuned reasoning model}. All tests are repeated 3 times on different seeds on two identical NVIDIA L40S GPUs. We run experiments with the following backbone models\footnote{\textit{Model suffixes (\eg, ``Reasoning'', ``Instruct'') are omitted in tables for brevity.}}:
\begin{inparaenum}[(i)]
    \item Foundation-Sec-8B-Reasoning~\cite{weerawardhena2025fndtsec},
    \item Ministral-3-8B-Reasoning~\cite{liu2026ministral},
    \item Ministral-3-3B-Instruct~\cite{liu2026ministral},
    \item Qwen3-8B~\cite{yang2025qwen3}.
\end{inparaenum}

\textbf{Metrics.}
We evaluate the quality of extracted knowledge graphs along three complementary dimensions:

\textit{Triplet Extraction.}
Given a set of gold triplets \( \mathcal{G} \) and predicted triplets \( \mathcal{P} \), we adopt a \emph{semantic matching}~\cite{ge2023knowledge, zhu2017sematch}. Each triplet \( t = (s, r, o) \) is encoded into a dense vector representation \( \phi(t) \) using a sentence embedding model (\eg, all-mpnet-base-v2\footnote{\href{https://huggingface.co/sentence-transformers/all-mpnet-base-v2}{https://huggingface.co/sentence-transformers/all-mpnet-base-v2}}) applied to the concatenation of its components. Pairwise similarities are computed using cosine similarity~\cite{lairgi2026atom, bouchiha2024llmchain} with a threshold, $T=0.6$ (see the Appendix~\ref{app:sim_thr} and~\ref{app:semantic_eval} for ablation). We then compute an optimal one-to-one alignment between \( \mathcal{G} \) and \( \mathcal{P} \). Based on this alignment, we compute Precision (\%), Recall (\%), and F1-score (\%):
\[
\text{Precision} = \frac{TP}{TP + FP}, 
\text{Recall} = \frac{TP}{TP + FN}, 
\text{F1} = \frac{2 \cdot TP}{2 \cdot TP + FP + FN}.
\]

\textit{Entity Typing.}
We evaluate typing quality at two levels. First, we report \emph{Partial Typing Accuracy (PTA)}, which corresponds to the standard entity typing (or classification) accuracy commonly used in prior work~\cite{cheng2025ctinexus, cheng2025crucialg}. PTA evaluates entity types independently and assigns credit when only one of the two (subject or object) is correctly predicted:
\[
\text{PTA} = \frac{N_{\text{correct}} + 0.5 \cdot N_{\text{partial}}}{N}.
\]
Second, we introduce a stricter metric, \emph{Full Typing Accuracy (FTA)}, which requires both subject and object types of a triplet to be correctly predicted. This provides a more holistic assessment of typing quality at the triplet level.

\textit{Graph Structural Similarity.}
To assess the overall structural quality of the predicted graph, we compute a semantic variant of the Jaccard similarity between the sets of gold and predicted triplets. Let \( M \) denote the number of semantically matched triplets under threshold \( T \). The graph similarity (\%) is defined as:
\[
\text{GraphSim} = \frac{M}{|\mathcal{G}| + |\mathcal{P}| - M}.
\]
This metric captures both structural overlap and semantic equivalence between the predicted and gold KGs. All metrics are reported as mean values (\%) over three runs (max std $\leq$ 1.0\%). \\

\textit{Computational Efficiency.} We evaluate CSKG construction time. Detailed results are provided in Appendix~\ref{app:time}.

\subsection{RQ1: Does agentic decomposition improve extraction quality?}

\begin{table*}[t]
\centering

\begin{minipage}{0.48\textwidth}
\centering
\caption{\small Best performance comparison on TEST0, TEST1 and TEST2.}
\label{tab:best_results}
\resizebox{\columnwidth}{!}{
\Large
\begin{tabular}{l l cc cc cc}
\toprule
\multirow{2}{*}{Category} & \multirow{2}{*}{Model(s)} 
& \multicolumn{2}{c}{TEST0} 
& \multicolumn{2}{c}{TEST1} 
& \multicolumn{2}{c}{TEST2} \\
\cmidrule(lr){3-4} \cmidrule(lr){5-6}
 & & F1 & Graph & F1 & Graph & F1 & Graph \\
\midrule
\multirow{2}{*}{Best CTINEXUS} 
& DeepSeek-V3.1 
& 74 & 56
&72 &58 
&66 &52 \\
\cmidrule(lr){2-8}
& Devstral-2 
& 73 & 60 
&72 &58 
&71 &58 \\
\midrule
\multirow{2}{*}{Best TACTIC-KG}
& Ministral-3-3B 
& 73 & 62
&79 &63 
&76 &61 \\
\cmidrule(lr){2-8}

& Ministral-3-8B
& \textbf{77} & \textbf{62}
& \textbf{80} & \textbf{67} 
& \textbf{78} & \textbf{63} \\
\bottomrule
\end{tabular}}
\end{minipage}
\hfill
\begin{minipage}{0.48\textwidth}
\centering
\caption{\small TACTIC-KG performance (Model + LoRA) on TEST1.}
\label{tab:TACTIC-KG_ministral}
\resizebox{\columnwidth}{!}{
\Large
\begin{tabular}{llcccccc}
\toprule
Model & Role & Precision & Recall & F1 & FTA & PTA & Graph \\
\midrule
\multirow{3}{*}{Ministral-3-3B~\cite{liu2026ministral}} 
& Ext./Typ. &80 &79 & \textbf{79} &51 &74 & \textbf{63} \\
& Verifier  & \textbf{83} &68 &74 & \textbf{52} & \textbf{75} &58 \\
& Curator   &76 & \textbf{77} &75 &51 &74 &61 \\
\midrule
\multirow{3}{*}{Ministral-3-8B~\cite{liu2026ministral}} 
& Ext./Typ. &73 & \textbf{90} & \textbf{80} & \textbf{52} & \textbf{72} &67 \\
& Verifier  & \textbf{78} &85 & \textbf{80} &50 &71 & \textbf{67} \\
& Curator   &71 & \textbf{90} &78 &48 &70 &63 \\
\bottomrule
\end{tabular}}
\end{minipage}

\end{table*}

Table~\ref{tab:best_results} compares TACTIC-KG with monolithic baselines.

\textit{Observation 1: Higher overall performance.}  
TACTIC-KG achieves up to \textbf{80} F1 and \textbf{67} GraphSim with Ministral-8B, outperforming large monolithic models such as DeepSeek-V3.1 (72 / 58).

\textit{Observation 2: Significant recall improvement.}  
Agent-based extractors reach up to \textbf{90} recall, compared to $\leq$ 66 for monolithic baselines (Table~\ref{tab:ctinexus_results}).

\textit{Observation 3: Limitations of ICL for smaller models.}  
ICL benefits large models, but smaller ones struggle (\eg, GPT-OSS-20B reaches 47 F1), which proves the limited generalization from prompts (Sect.\ref{subsec:context-size}).

\textit{Answer to \textbf{RQ1.}} Agentic decomposition with parameter specialization via LoRA significantly improves extraction quality by increasing recall and overall F1, enabling smaller models to match or outperform much larger ICL systems.

\subsection{RQ2: How does agent specialization affect PR trade-offs?}

TACTIC-KG exhibits distinct behaviors across agent roles (Table~\ref{tab:TACTIC-KG_ministral}).

\textit{Observation 1: Extractors favor recall.}  
Extractors may aggressively explore candidate relations, achieving high recall (up to \textbf{90}) at moderate precision.

\textit{Observation 2: Verifiers balance precision and recall.}  
Verifiers filter hallucinations and improve F1 (up to \textbf{80}) while maintaining strong recall.

\textit{Observation 3: Curators improve completeness.}  
Curators recover missing links and further increase recall (up to \textbf{85}), improving graph completeness.

\textit{Answer to \textbf{RQ2.}}  
Agent specialization enables explicit control of the precision--recall trade-off by distributing complementary roles across stages, achieving high recall while maintaining stable precision. In contrast, ICL monolithic models rely on single-pass reasoning, which struggles to reliably infer unstated relations. This often leads to over-completion or artificial connectivity, where spurious links are introduced between subgraph nodes and the main topic node (Sect.~\ref{subsec:overcopletion}).

\begin{table*}[t]
\centering

\begin{minipage}{0.48\textwidth}
\centering
\caption{\small CTINEXUS baselines (Model + ICL) on TEST1.}
\label{tab:ctinexus_results}
\resizebox{\columnwidth}{!}{
\Large
\begin{tabular}{lcccccc}
\toprule
Model & Precision & Recall & F1 & FTA & PTA & Graph \\
\midrule
GPT-OSS (20B)~\cite{agarwal2025gpt} &64 &40 &47 &28 &48 &36 \\
Nemotron-3 Nano (30B)~\cite{blakeman2025nemotron} &64 &57 &58 &16 &44 &44 \\
GPT-OSS (120B)~\cite{agarwal2025gpt} &76 &56 &62 &43 &67 &48 \\
GLM-5~\cite{zeng2026glm} &71 &61 &65 &46 &69 &50 \\
Kimi-K2 (1T)~\cite{team2025kimi} &82 &66 &72 &47 &71 &57 \\
Devstral-2 (123B)~\cite{rastogi2025devstral} & \textbf{83} &65 &72 & \textbf{50} & \textbf{72} &58 \\
DeepSeek-V3.1 (671B)~\cite{deepseekai2024deepseekv3technicalreport} & \textbf{83} & \textbf{66} & \textbf{72} &45 &69 & \textbf{58} \\
\bottomrule
\end{tabular}}
\end{minipage}
\hfill
\begin{minipage}{0.48\textwidth}
\centering
\caption{\small TACTIC-KG performance with hybrid models on TEST1.}
\label{tab:TACTIC-KG_hybrid}
\resizebox{\columnwidth}{!}{
\Large
\begin{tabular}{llcccccc}
\toprule
Model & Role & Precision & Recall & F1 & FTA & PTA & Graph \\
\midrule

\multirow{3}{*}{Foundation-Sec-8B~\cite{weerawardhena2025fndtsec}} 
& Ext./Typ. &81 &76 &77 &52 &74 &63 \\
& Verifier  & \textbf{86} &73 & \textbf{78} & \textbf{53} & \textbf{75} &63 \\
& Curator     &77 & \textbf{80} &78 &48 &72 & \textbf{64} \\

\midrule

\multirow{3}{*}{Ministral-3-3B~\cite{liu2026ministral}} 
& Ext./Typ. &80 &79 &79 & \textbf{51} & \textbf{74} &63 \\
& Verifier  & \textbf{82} &75 &78 &50 &73 &63 \\
& Curator     &70 & \textbf{83} & \textbf{75} &47 &72 & \textbf{69} \\
\midrule

\multirow{3}{*}{Qwen3-8B~\cite{yang2025qwen3}} 
& Ext./Typ. &84 &75 &78 &50 &71 &68 \\
& Verifier  & \textbf{89} &74 & \textbf{80} & \textbf{52} & \textbf{72} & \textbf{67} \\
& Curator    &77 & \textbf{85} & \textbf{80} &50 & \textbf{72} &66 \\

\bottomrule
\end{tabular}
}

\end{minipage}

\end{table*}

\subsection{RQ3: Does TACTIC-KG improve typing, semantic grounding, and structural consistency?}

TACTIC-KG demonstrates consistent improvements in typing and semantic grounding (Tables~\ref{tab:TACTIC-KG_ministral} and~\ref{tab:ctinexus_results}).

\textit{Observation 1: Higher typing accuracy.}  
TACTIC-KG achieves up to \textbf{72} PTA and \textbf{52} FTA, outperforming strong ICL baselines such as DeepSeek ($\approx$69 PTA).

\textit{Observation 2: Improved semantic grounding.} 
Reduced cognitive load (LoRa vs. ICL demos) on individual agents and cleaner intermediate representations with ontology-aware verification reduce hallucinations and improve typing.

\textit{Observation 3: Better structural consistency.}  
Graph scores reach up to \textbf{68}, compared to 58 for DeepSeek, indicating more coherent graph construction.

\textit{Answer to \textbf{RQ3.}} 
TACTIC-KG improves typing accuracy, semantic grounding, and structural consistency by enforcing ontology-aware reasoning and reducing error propagation across stages (Sect.~\ref{subsec:faithfulness}). The observed gains are a direct result of the reduced load and the constrained verification and curation (\ie, the ontology is used during both fine-tuning and inference). Thus, any new entity or relation must remain consistent with ontology schema, which in turn limits type ambiguity and reduces hallucinated links.

\subsection{RQ4: What is the impact of a hybrid agentic design?}

Table~\ref{tab:TACTIC-KG_hybrid} shows that TACTIC-KG improves performance with shared reasoning (Ministral-8B + LoRA for verification and curation).

\textit{Observation 1: Improved stability across roles.}  
Hybrid configurations maintain consistently high precision, recall, and graph scores across all roles.

\textit{Observation 2: Enhanced reasoning consistency.}  
Shared reasoning models improve typing and partial typing (up to \textbf{72} PTA), while stabilizing outputs.

\textit{Observation 3: Balanced performance.}  
For example, Qwen3 achieves strong precision (\textbf{89}), recall (\textbf{85}), and graph scores (\textbf{67}).

\textit{Answer to \textbf{RQ4.}} 
Hybrid agentic design improves robustness and consistency by decoupling \emph{generation} from \emph{validation}. Using a stronger, shared reasoning model for verification and curation allows consistent filtering over heterogeneous extractor outputs. This model enforces ontology constraints and corrects local extraction errors. As a result, different extractors can either prioritize recall or precision, while the centralized verifier--curator pair suppresses hallucinations and resolves conflicts, leading to more stable and higher-quality final graphs.

\begin{table*}[t]
\centering
\small
\caption{Impact of training data volume and diversity on TACTIC-KG (Extractor \& Typer) performance on TEST2.} 
\label{tab:rq5-data-scaling}
\setlength{\tabcolsep}{4pt}
\renewcommand{\arraystretch}{1.2}
\resizebox{0.8\columnwidth}{!}{
\begin{tabular}{l l c c c c c c}
\toprule
\Large
\textbf{Version} & \textbf{Model} & \textbf{Precision} & \textbf{Recall} & \textbf{F1} & \textbf{FTA} & \textbf{PTA} & \textbf{Graph} \\
\midrule
\multirow{2}{*}{TC-1: Diverse, 539 chunks} 
& Ministral-3-8B &72 & \textbf{89} & \textbf{78} & \textbf{48} & \textbf{70} & \textbf{63} \\
& Ministral-3-3B  &71 &85 &76 &42 &66 &61 \\
\midrule
\multirow{2}{*}{TC-2: Single, 201 chunks } 
& Ministral-3-8B &73 &85 &76 &47 &68 &60 \\
& Ministral-3-3B   &78 &77 &76 &48 &70 &61 \\
\midrule
TC-3: Single, 100 chunks 
& Ministral-3-3B &69 &79 &72 &44 &68 &58 \\
\midrule
TC-4: Diverse, 192 chunks 
& Ministral-3-3B & \textbf{79} &79 & \textbf{78} &47 &70 & \textbf{63} \\
\midrule
\multirow{3}{*}{CTINEXUS: Models + ICL}
& Devstral-2:123b &80 &68 &72 &48 &71 &58 \\
& Deepseek-v3.1:671b & \textbf{82} &57 &66 &53 &74 &52 \\
& Kimi-k2:1t &80 &63 &69 &47 &68 &56 \\
\bottomrule
\end{tabular}
}
\end{table*}

\subsection{RQ5: How much training data does TACTIC-KG need to reach monolithic model performance?}

Table~\ref{tab:rq5-data-scaling} reports the performance of TACTIC-KG under different training confi\-gurations (TC) and model sizes on TEST2. We compare \begin{inparaenum}[(i)] 
\item TC-1, where models are trained on a larger and more diverse dataset (185 reports from \cite{della2025cti} + \cite{cheng2025ctinexus}), 
\item TC-2, where the extractor and the typer are fine-tuned on 108 reports from \cite{cheng2025ctinexus} data only,
\item TC-3, where the training is restricted to a fraction of \cite{cheng2025ctinexus} only and,
\item TC-4, where the training uses a fraction of both datasets.\end{inparaenum}

\textit{Observation 1: Performance improves with more data.}  
Increasing training data from 100 to 201 chunks improves F1 (72 $\rightarrow$ 76) and recall (79 $\rightarrow$ 85).

\textit{Observation 2: Diversity matters more than volume.}  
Using more diverse data (TC-1, TC-4) yields the best performance (F1 \textbf{78}, PTA \textbf{70}).

\textit{Answer to \textbf{RQ5.}}  TACTIC-KG requires only moderate training data ($\sim$200 chunks) to reach strong performance, with data diversity playing a more critical role than volume.

\section{Discussion}
\label{sec:discussion}
Results demonstrate that agent specialization via LoRA improves both stability and extraction quality, even when using smaller models. TACTIC-KG consistently outperforms larger ICL monolithic pipelines across F1, recall, and graph quality, highlighting the benefits of task decomposition. By separating roles, the framework enables finer control over the precision--recall trade-off, allowing extractors to prioritize coverage while downstream stages refine correctness. Graph consistency is further enhanced through verification and curation, which reduce error propagation and enforce coherence. Moreover, a hybrid design combining specialized extractors and typers with a shared reasoning model (verifier and curator) provides additional gains in overall performance. Despite these improvements, full typing accuracy remains a bottleneck across all systems.

\section{Conclusion}
\label{sec:conclusion}
This paper introduced TACTIC-KG, an agentic framework for constructing Cyber Threat Intelligence Knowledge Graphs. By decomposing CSKG construction into specialized small LLM agents, TACTIC-KG achieves higher accuracy, stability, and cost-efficiency than larger monolithic LLM-based approaches. Although this work focuses on publicly available CTI reports, the proposed framework naturally extends to enterprise settings, enabling organizations to build and maintain CSKGs while keeping sensitive data local, with minimal deployment overhead. \\
Future work will focus on robust graph completion, stronger ontology constraints, and CSKG repair under noisy and adversarial CTI inputs.

\medskip

\small \textbf{Acknowledgments.} This research is supported by the CKRISP project (ANR-23-IAS4-0001).

\appendix

\begin{table}[th]
\begin{minipage}{0.48\textwidth}
\centering
\caption{Impact of similarity threshold on entity and relation merging (embedding model: \textit{all-MiniLM-L12-v2}).}
\label{tab:threshold_analysis}
\resizebox{\columnwidth}{!}{
\begin{tabular}{ccccccc}
\hline
\multirow{2}{*}{$T$} 
& \multicolumn{3}{c}{\textbf{Verifier}} 
& \multicolumn{3}{c}{\textbf{Curator}} \\
& Precision & Recall & F1 
& Precision & Recall & F1 \\
\hline

0.6 &83 &66 &73 & \textbf{73} &79 &75 \\
0.7 & \textbf{84} &72 &77 &72 &80 &75 \\
0.8 &81 &75 &78 &69 &84 &75 \\
0.9 &81 & \textbf{77} & \textbf{78} &70 & \textbf{84} & \textbf{76} \\

\hline
\end{tabular}}
\end{minipage}
\begin{minipage}{0.48\textwidth}
\centering
\small
\caption{The impact of the evaluation embedding models on TACTIC-KG (Ministral-3B).}
\label{tab:embedding_ablation}
\resizebox{\columnwidth}{!}{
\begin{tabular}{lccc|ccc|ccc}
\toprule
\multirow{3}{*}{Embedding Model (T)}
 & \multicolumn{3}{c}{Extraction}
 & \multicolumn{3}{c}{Verification}
 & \multicolumn{3}{c}{Curation} \\
\cmidrule(lr){2-4}\cmidrule(lr){5-7}\cmidrule(lr){8-10}
 & P & R & F1 & P & R & F1 & P & R & F1 \\
\midrule
MiniLM-L6 (0.6) &77 &77 &76 &79 &75 &76 &70 &83 &75 \\
MPNet (0.6)    &79 &78 &78 &79 &75 &76 &70 &84 &76 \\
MiniLM-L12 (0.6) &80 &79 &79 &81 &77 &78 &70 &84 &76 \\
MiniLM-L12 (0.7) &69 & 69 & 69 & 71 & 65 & 68 &  62 & 73 & 66 \\
Paraphrase (0.6) & \textbf{81} & \textbf{81} & \textbf{80} & \textbf{82} & \textbf{78} & \textbf{80} & \textbf{72} & \textbf{86} & \textbf{77} \\
\bottomrule
\end{tabular}}
\end{minipage}
\end{table}

\section{Impact of Similarity Threshold on Entity/Relation Merging.} \label{app:sim_thr}
We analyze the effect of the similarity threshold $T \in \{0.6, 0.7, 0.8, 0.9\}$ used for clustering semantically similar entities and relations (Table~\ref{tab:threshold_analysis}). This threshold controls the granularity of canonicalization: lower values encourage aggressive merging, while higher values enforce stricter matching. Increasing $T$ consistently improves recall (\eg, curated recall rises from 79 to 84), as stricter merging preserves distinct semantics. Precision shows a non-monotonic trend—peaking at moderate thresholds and slightly decreasing at higher $T$ due to under-merging. Overall, F1 improves with higher $T$ (up to 78 in verification and 76 in curation), indicating a better precision–recall balance.

\begin{table}[t]

\small
\begin{minipage}{0.44\textwidth}
\centering
\caption{Impact of the evaluation embedding models on CTINEXUS (DeepSeek).}
\label{tab:ctinexus_embedding}
\resizebox{0.95\columnwidth}{!}{
\begin{tabular}{lccc}
\toprule
Embedding Model (T) & P & R & F1 \\
\midrule

all-MiniLM-L6-v2 (0.6)
&78 &62 &67 \\

all-mpnet-base-v2 (0.6) 
&83 &66 &72 \\
all-MiniLM-L12-v2 (0.6)
&83 &66 &72 \\
all-MiniLM-L12-v2 (0.7) & 69 & 54 & 59 \\
paraphrase-MiniLM-L6-v2 (0.6)
& \textbf{83} & \textbf{67} & \textbf{72} \\

\bottomrule
\end{tabular}}
\end{minipage}
\begin{minipage}{0.5\textwidth}
\centering
\caption{Per-stage and end-to-end time overhead of models (seconds).}
\label{tab:model_timings}
\resizebox{\columnwidth}{!}{
\begin{tabular}{lccccc}
\toprule
Model & IE & ET & VER/EA & CUR/LP & E2E \\
\midrule
\multicolumn{6}{c}{\textbf{CTINEXUS}} \\
Kimi-k2:1t & 9 & 14 & 0-1 & 0-1 & 23 \\
Deepseek-v3.1 & 22 & 47 & 0-1 & 0-1 & 72 \\
Devstral-2:123b & \textbf{5-6} & \textbf{6-7} & \textbf{0-1} & \textbf{0-1} & \textbf{12} \\
\midrule
\multicolumn{6}{c}{\textbf{TACTIC-KG}} \\
Foundation-Sec-8B & 31 & 53 & 58 & 18 & 155 \\
Qwen3-8B & 46 & 183 & 59 & 17 & 297 \\
Ministral-3-8B & 26 & 26 & 67 & 25 & 145 \\
Ministral-3-3B & \textbf{12} & \textbf{23} & \textbf{58} & \textbf{21} & \textbf{115} \\
\bottomrule
\end{tabular}}
\end{minipage}
\end{table}

\section{Impact of Embedding Models on Semantic Evaluation} \label{app:semantic_eval}
Tables~\ref{tab:embedding_ablation} and~\ref{tab:ctinexus_embedding} evaluate the effect of the embedding model used to compute semantic similarity between predicted and gold triplets ($T\in[0.6, 0.7]$). For TACTIC-KG, the extraction (Ministral-3B) and verification (Ministral-8B) models remain fixed; only the similarity function varies. Across both our pipeline and CTINEXUS (with \textit{Deepseek}), results show a consistent performance gap between embedding models. \textit{paraphrase-MiniLM-L6-v2} achieves the best overall performance, with gains of up to +1.5 F1, while \textit{all-MiniLM-L6-v2} yields the lowest scores, due to limited representational capacity. Variations are primarily driven by recall rather than precision: precision remains relatively stable, whereas recall increases with paraphrase embeddings (\eg, from 83 to 86), which means that more predictions are considered semantically correct. In contrast, \textit{all-mpnet-base-v2} consistently produces slightly lower recall and F1 with a stricter semantic matching criterion. This trend holds across both pipelines, which highlights the importance of reporting results with multiple embedding models.

\section{Execution Time and E2E Analysis} \label{app:time}
Table~\ref{tab:model_timings} reports per-stage and end-to-end (E2E) execution times. We note that these timings are indicative only: CTINEXUS pipline was executed on the OLLAMA cloud, whereas TACTIC-KG models ran locally on NVIDIA L40S GPUs. Differences in hardware, concurrency, and runtime configurations can therefore impact direct comparisons. CTINEXUS large models (\eg, Deepseek-v3.1) achieve fast inference for individual tasks. Specialized agents demonstrate higher per-stage costs for reasoning and verification due to their multi-step design, but they benefit from improved extraction quality and graph consistency. Lightweight or smaller models (\eg, Ministral-3B) achieve competitive times with robust construction performance.

\bibliographystyle{splncs04}
\bibliography{refs}

\end{document}